\documentclass[final,1p,times]{elsarticle}

\usepackage{amsmath,amssymb,amsfonts,amscd}
\usepackage{graphicx}

\usepackage{color,soul}

\begin{document}

\begin{frontmatter}

\title{Quantum phenomenology of conjunction fallacy}
\author[label1]
{Taksu Cheon \corref{cor1}}
\ead{taksu.cheon@kochi-tech.ac.jp}
\author[label2,label3]
{Taiki Takahashi}
\ead{ttakahashi@lynx.let.hokudai.ac.jp}
\address
[label1]
{Laboratory of Physics, Kochi University of Technology,
Tosa Yamada, Kochi 782-8502, Japan}
\address
[label2]
{
Department of Behavioral Science, Faculty of Letters,
Hokkaido University,
Kita-ku, Sapporo, 060-0810, Japan}
\address
[label3]
{
Center for Experimental Research in Social Sciences, 
Hokkaido University,
Kita-ku, Sapporo, 060-0810, Japan}

%
\cortext[cor1]{corresponding author}

\date{\Today}

\begin{abstract}
A quantum-like description of human decision process is developed, 
and a heuristic argument supporting the theory as sound phenomenology is given.  
It is shown to be capable of quantitatively explaining the conjunction fallacy
in the same footing as the violation of sure-thing principle.
\end{abstract}


\end{frontmatter}



\section{Introduction}
Quantum game theory, which is the  description of decision process of two quantum agents mutually dependent on other's decision, has had qualified success in analyzing some of the puzzling problems in game theory \cite{ME99, EW99,CT06}.  
When one of the agent in the game theory is ``frozen'', namely, the decision of one of the agent is pre-fixed, it becomes a two stage decision process for a single quantum agent, and this is the setting in which the quantum decision theory emerges \cite{AA94, YS08, PB09, KH09, FR09, CT10, BK11, BP11}.
The source of the phenomenological success of the quantum game theory
in non-quantum settings is the flexibility of the theory
capable of effectively incorporating the altruism of the player through the language of quantum entanglement.  
Quantum decision theory also has seen some successes in describing two long-standing
psychological paradoxes, the violation of sure-thing principle \cite{TS92, ST92}, and the conjunction fallacy \cite{TK83, SS90}.
It is very desirable, therefore, to look into the detailed workings of the quantum decision theory, 
to examine the phenomenological effectiveness in more systematic fashion, and to identify
its physical contents.
%

The purpose of this paper is two-fold.
We first examine the relation between sure-thing-principle violation and the conjunction fallacy.
Next, we examine the ``quantum'' formula of decision under unknown condition 
in terms of arithmetic and geometric averages, and try to make sense as a useful phenomenology. 
The experimental numbers of conjunction fallacy are analyzed with the quantum formula, and a satisfactory quantitative success is achieved.

\section{Conjunction fallacy and violation of sure-thing principle}

Consider an agent who faces a two stage probabilistic process made up a condition and the decision.
Condition is chosen from two possibilities, which we respectively call condition $0$ and $1$, to which we assign probabilities $q_0$ and $q_1$. 
The decision of the agent consists of choosing from two events, choices $0$ and $1$, 
to which we assign two numbers $p_{0j}$ and $p_{1j}$, for each $j=0$, and $1$,  
which respectively  signify the probabilities of the agent choosing $0$ and $1$ under the condition $j$.

Let us now ask what the probability  is of the agent making choice $k$ irrespective of the condition.
The law of total probability tells that this probability, which we call $Q_k$, has to be 
the weighted sum of conditional probabilities  $p_{k0}$ and $p_{10}$ with respective weights $q_0$ and $q_1$;
\begin{eqnarray}
\label{ee01}
Q_k = p_{k0} q_0 + p_{k1} q_1 .
\end{eqnarray}
Natrurally, all probablistic quantities $q_j$ and $p_{kj}$ are non-negative real numbers.
In particular, we have
\begin{eqnarray}
\label{ee02}
q_0 \ge 0, \quad q_1 \ge 0,
\end{eqnarray}
from which we easily obtain
\begin{eqnarray}
\label{ee03}
{\rm Max}(p_{k0}, p_{k1}) \ge Q_k \ge {\rm Min}(p_{k0}, p_{k1}),
\end{eqnarray}
namely, $Q_k$ has to be a number in between $p_{k0}$ and $p_{k1}$, which is often referred
to as the {\it sure-thing principle}.

From another trivial relations
\begin{eqnarray}
\label{ee04}
p_{k0} q_0 \ge 0, \quad 
p_{k1} q_1 \ge 0,
\end{eqnarray}
we obtain
\begin{eqnarray}
\label{ee05}
Q_k \ge p_{k0} q_0,\quad
Q_k \ge p_{k1} q_1 ,
\end{eqnarray}
%
which simply express the logical inclusivity between the unconditional
event $k$ (a ``constituent'') whose probability
is $Q_k$, and the conditional component events (``conjunctions'') whose
probabilities are given by $p_{k0} q_0$ and $p_{k1} q_1$.

The experiments by Tversky and Shafir show that the sure-thing principle, (\ref{ee03}), can be broken
in a two stage prisonerÕs dilemma game \cite{TS92, ST92}.  
Separately, Tversky and Kahneman \cite{TK83} have found that the human often make
seemingly illogical judgement that
contradicts with logical inclusivity (\ref{ee05}), 
and judge that an instance is more likely to be a member of a conjunctive category than of one of its constituent categories, to which they coin the term {\it conjunction fallacy}. 
One vivid example is given by the story on Linda.  People are presented with descriptions of a bright young woman  who was a philosophy major and concerned with social justice.  Subsequent answers to questionaires reveal that people believes it more likely that she is a {\it feminist bankteller} than just a bankteller,
indicating the violation of inequalities (\ref{ee05}) for subjective probabilities involving human judgement.

\section{Quantum-like description}
Human sense often responds to the stimulus logarithmically rather than linearly.   
A recent study indicates that human subjective probabilities have nonlinear psychophysical 
foundations \cite{TA11}.
In such cases,
average response to two evens would be given by the geometric average, not the algorithmic average.
If such is the case for perception of two conjunction probabilities $p_{k0} q_0$ and $p_{k1} q_1$, we will perceive
\begin{eqnarray}
\label{ee11}
Q^{G}_k = 2\sqrt{ p_{k0} q_0 p_{k1} q_1} ,
\end{eqnarray}
as the average of two probabilities, rather than the probability given by (\ref{ee01}).  It is possible that the cognitive process sometime involves both
elements (\ref{ee01}) and (\ref{ee11}), and is given by their linear superposition
\begin{eqnarray}
\label{ee12}
Q_k = N_1 \cdot \left( p_{k0} q_0 + p_{k1} q_1 \right) + N_2 \cdot 2 \sqrt{ p_{k0} q_0 p_{k1} q_1} .
\end{eqnarray}
Since the geometric average can be only as big as the arithmetic average, $N_2$ must be smaller than $N_1$ in order to ensure
the non-negativity of  this expression for all values of $q_j$ and $p_{kj}$.  This observation leads to
a parametrization using trigonometric functions with phase parameters $\theta_k$,
\begin{eqnarray}
\label{ee13}
Q_k = N \left(p_{k0} q_0 + p_{k1} q_1 
+ 2 \sqrt{ p_{k0} q_0 p_{k1} q_1} \cos\theta_k \right) .
\end{eqnarray}
Finally, from the requirement of the unity of total probability, $Q_0+Q_1 = 1$, $N$ can be determined.
Thus we obtain the explicit form for the generalized average probability for two stage decision process;
\begin{eqnarray}
\label{ee14}
Q_k = \frac{ p_{k0} q_0 + p_{k1} q_1  + 2 \sqrt{ p_{k0} q_0 p_{k1} q_1} \cos\theta_k }
{ 1  + 2 \sqrt{ p_{00} q_0 p_{01} q_1} \cos\theta_0 + 2 \sqrt{ p_{10} q_0 p_{11} q_1} \cos\theta_1 } .
\end{eqnarray}
This two parameter expression happens to coincide exactly with the result obtained from the {\it quantum decision theory} \cite{CT10}.
It is easy to show that, for unknown $\theta_k$, this expression is bounded by
\begin{eqnarray}
\label{ee15}
\frac{ ( \sqrt{p_{k0} q_0} + \sqrt{p_{k1} q_1} )^2 }
{ 1  +2 \sqrt{ q_0 q_1} (\sqrt{ p_{k0}  p_{k1} }-  \sqrt{ p_{{\bar k}0}  p_{{\bar k}1} } ) } 
\ge Q_k \ge 
\frac{ ( \sqrt{p_{k0} q_0} - \sqrt{p_{k1} q_1} )^2 }
{ 1  - 2 \sqrt{ q_0 q_1} (\sqrt{ p_{k0}  p_{k1} }-  \sqrt{ p_{{\bar k}0}  p_{{\bar k}1} } ) } .
\end{eqnarray}
For unknown $q_j$,  on top of indeterminate $\theta_k$, we obtain the trivial inequality
\begin{eqnarray}
\label{ee16}
1 \ge Q_k  \ge 0 .
\end{eqnarray}
It is possible, with proper values of $\theta_k$ to break either of the two ``classical'' 
ineualities (\ref{ee03}) and (\ref{ee05}) or the both.

It is possible to conjecture that there is a way to construct a neural model that processes two stage probabilistic event and produces 
this type of result, which could be a good simulation of the actual neural network at work in our brain.
In fact, a recent neuroeconomic study
demonstrated that nonlinearity in subjective probabilities is
determined with dopamine D1 receptors in the human brain \cite{TM10}.
If, for some reason, this setup with specific value of $\theta_k$ had proven  evolutionarily advantageous, 
as a mean to achieve optimal risk management against unknown conditions of our environment, for example, it would be likely that such seemingly ``illogical'' traits have been left in our mind.

\section{Relation to single-phase approach}
The quantum formula for two-choice condition and two-choice decision, (\ref{ee14}), has two independent parameters, $\theta_0$ and $\theta_1$, and is not simple enough for our intuition to grasp its meaning at the first glance.  We show that it can be reduced to one-parameter formula if some conditions are met for 
the probabilities and the parameters appearing there.  

Let us suppose that we can impose a parametric relation 
\begin{eqnarray}
\label{ee21}
 2 \sqrt{ p_{00} p_{01} } \cos\theta_0 + 2 \sqrt{ p_{10}  p_{11} } \cos\theta_1 = 0
\end{eqnarray}
on $\theta_0$ and $\theta_1$.  The expression (\ref{ee14}) is reduced to a very simple form
\begin{eqnarray}
\label{ee22}
Q_k = p_{k0} q_0 + p_{k1} q_1  + 2 \sqrt{ p_{k0} q_0 p_{k1} q_1} \cos\theta_k  .
\end{eqnarray}
If we further assume the condition
\begin{eqnarray}
\label{ee23}
p_{00} p_{01}=p_{10} p_{11} ,
\end{eqnarray}
we obtain 
$\cos\theta_0 = -\cos\theta_1$, and with the notation $\theta = \theta_0$, we have
\begin{eqnarray}
\label{ee24}
Q_k = p_{k0} q_0 + p_{k1} q_1  +(-1)^k \,2 \sqrt{ p_{k0} q_0 p_{k1} q_1} \cos\theta  ,
\end{eqnarray}
which is identical the Franco's approach \cite{FR09} for the analysis conjunction fallacy.  
It shoud be noted that we may not always be lucky enough to be able to assume (\ref{ee23}), since
probabilities $p_{kj}$ is a given input quantity for each experiment, not an adjustable parameter.

\section{Analysis of experimental numbers}
There is a set of fourteen pairs of data on conjunction fallacy readily available,
coming the experiments conducted by Shafir {\it et al.} \cite{SS90}.
We first pick up two of them to fix the phase parameters, and then see whether
the rest of the data are well-described by our formula (\ref{ee14}).
The data are incomplete in the sense that not all input probabilities are measured.
At first sight, therefore, 
it appears unlikely that we can determin the quantum parameters from the data as they are.  
However, we shall show in the following, that with certain plausible assumptions, 
the information on the phase can be extracted  if two data are used in combination.

We take up two numbers from the experiment (1-i) and (1-c) of the Table 1 of \cite{SS90},
that respectively typify ``low probability'' and ``high probability'' events.
Participants in the experiment are given the information (``instance'') ``Linda was a philosophy major.  She is bright and concerned with issue of discrimination and social justice.''

In the experiment (1-i),
the conditional event (``constituent'') is given as the answer to the question whether Linda is a feminist (1) or not (0).  The decision event (another ``constituent'') is given as the answer to the question whether Linda is a bank teller (1) or not (0). The probabilities $q_1$, $q_0$,  respectively corresponding to Linda being  a feminist or not, are not specified, but $q_1\equiv q$ is supposed to be ``a large number'', meaning $q>\frac{1}{2}$.  Experimental numbers for the conjunction probabilities under the condition (1) are given by  
\begin{eqnarray}
p_{11} q_1=0.401,
\quad 
p_{01} q_1 =q_1 - 0.401.
\end{eqnarray} 
The other conditional probabilities $p_{10}\equiv p$, $p_{00}=1-p$
are also not known, but $p_{10}$ is considered to be equal or bigger than $p_{11}$, namely 
$p_{10} \ge 0.401/q_1 $.
The probabilities under mixed condition are measured to be
\begin{eqnarray}
Q_1=0.241, \quad
Q_0 = 0.759.
\end{eqnarray}  

In the experiment (1-c),
the conditional event (``constituent'') is the same as in (1-i), and given as the answer to the question whether Linda is a feminist (1) or not (0).  The decision event (another ``constituent'') is given now as the answer to the question whether Linda is a teacher (1) or not (0). 
The conjunction probabilities under the condition (1) are experimentally determined to be   
\begin{eqnarray}
p_{11} q_1 = 0.601 ,
\quad p_{01} q_1 =q_1 - 0.601 .
\end{eqnarray}
For this case, we adopt the notation
$p_{10}\equiv p'$, $p_{00}=1-p'$ and assume $p_{10} \ge 0.601/q_1$.
The probabilities under mixed condition are measured to be
\begin{eqnarray}
Q_1=0.533, \quad Q_0 =  0.467.
\end{eqnarray}
%

If we assume that these two experimental values are described by single set of phase parameters
$\theta_j$, we have
\begin{eqnarray}
\label{ee25}
0.241 = \frac{ p (1-q) + 0.401  + 2 \sqrt{ 0.401 p (1-q)} \cos\theta_1 }
{ 1  + 2 \sqrt{ (q-0.401)(1-p) (1-q)} \cos\theta_0 + 2 \sqrt{ 0.401 p(1-q)} \cos\theta_1 } \, ,
\end{eqnarray}
\begin{eqnarray}
\label{ee26}
0.533 = \frac{ p' (1-q) + 0.601  + 2 \sqrt{ 0.601 p' (1-q)} \cos\theta_1 }
{ 1  + 2 \sqrt{ (q-0.601)(1-p') (1-q)} \cos\theta_0 + 2 \sqrt{ 0.601 p' (1-q)} \cos\theta_1 } \, .
\end{eqnarray}
We can revert this relation to obtain the phases $\theta_k$ as functions of $p$, $p'$, and $q$, namely,
\begin{eqnarray}
\label{ee27}
\theta_k =  \theta_k(p, p'; q) \, .
\end{eqnarray}
\begin{figure}[ht]
\center{
\includegraphics[width=6.5cm]{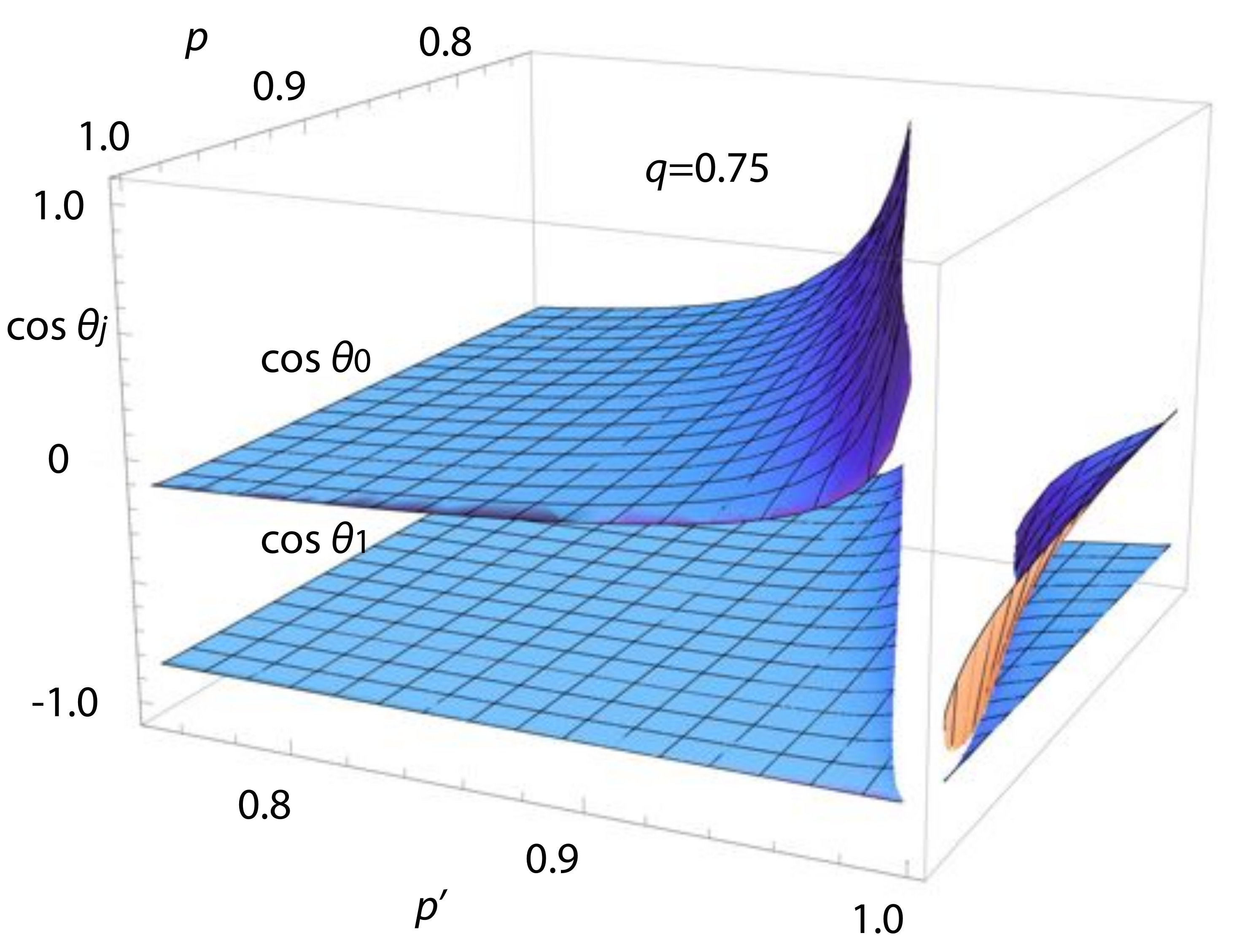}
}
\label{fig1}
\caption
{
This graph depicts the values of $\cos\theta_0$ and $\cos\theta_1$ for given values of $p$
and $p'$ with the assumption $q=0.75$.  
}
\end{figure}

In Figure.1, $\cos\theta_0$ and $\cos\theta_1$ are plotted as functions of $p=p_{11}$ for  (1-i),
and $p'=p_{11}$ fro (1-c) with the assumption $q=0.75$.
At the relevant region, $\cos\theta_j$ are more or less flat, and we have 
$\cos\theta_0=0.391$ and  $\cos\theta_1=-0.812$ at $p=0.601/q=0.801$, which gives the values
of the phases 
%
\begin{eqnarray}
\theta_0 = 0.52\pi , \quad  \theta_1 = 0.82\pi \quad ({\rm with \ } q_1=\frac{3}{4}) .
\end{eqnarray}
%

In Figure.2, we use this estimate to  plot theoretical prediction of $Q_1$ as a function of $p_{11}q_1$ along 
with all twenty eight experimental values taken from Table 1 of \cite{SS90}.  Here, $p_{10}$ is assumed to be equal to $p_{11}$.  The deviation from $Q_1(p_{11})= p_{11} q_1$, which is given by the dashed line, indicates the existence of the  conjunction fallacy.  From this figure, it is clear that the ``quantum'' prediction gives a rather good overall description of the experimental data that shows the existence of conjunction fallacy in statistically significant fashion.

Calculation with other values within the reasonable range  $1> q > 0.65$ reveals that 
$\theta_j(p, p', q)$ is a smooth and mild function of $q$ in this range, and the
situation essentially remains the same.
\begin{figure}[ht]
\center{
\includegraphics[width=6.0cm]{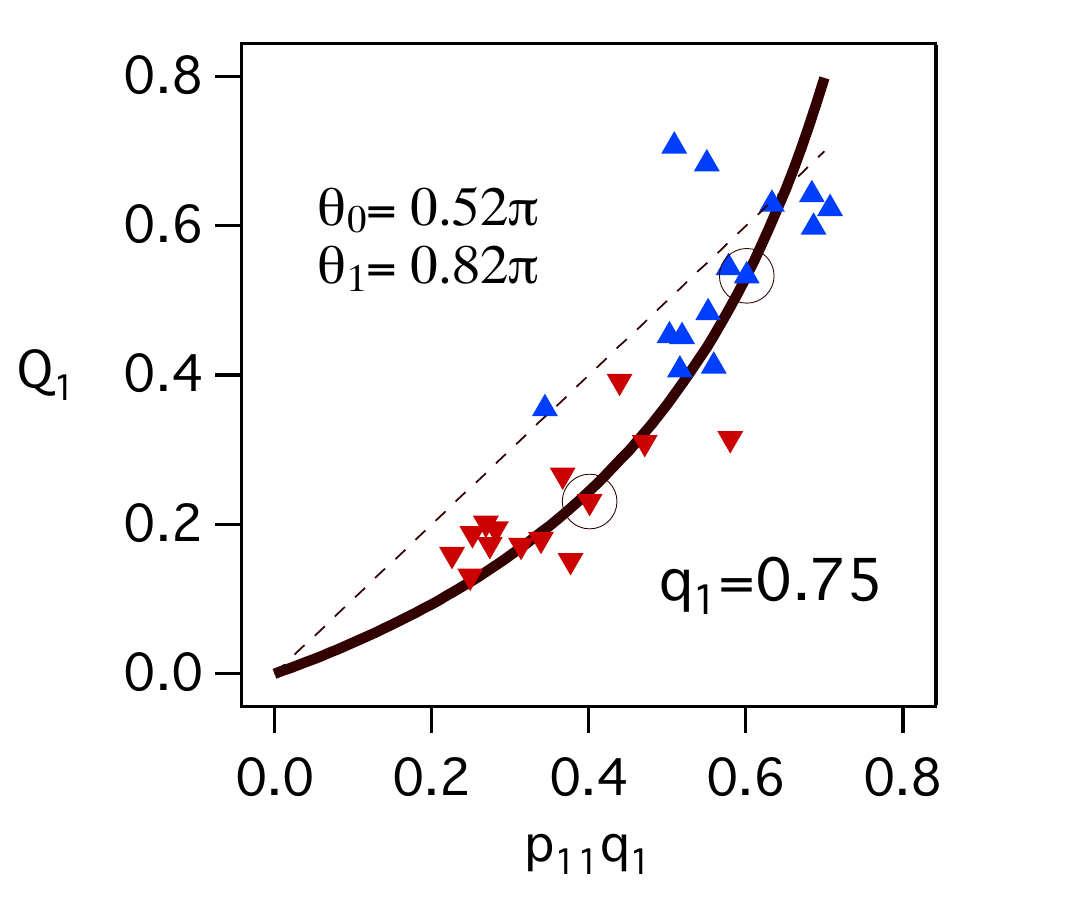}
}
\label{fig2}
\caption
{
This graph depicts theoretical prediction of  $Q_q$ versus $p_{11}q_1$ with the assumption  $q_1=0.75$
and the quantum phase parameters $\theta_0 = 0.52\pi$, $\theta_0 = 0.82\pi$.  Experimental data,
taken from \cite{SS90},
are shown with upward triangle for "(i)" series and downward triangle for "(c)" series.
Encircled data are the ones used to determined the phase parameter as shown in Figure.1.}
\end{figure}

When one of the phases, say $\theta_0$ is close to $\frac{\pi}{2}$, which is what we have essentially obtained in above estimate, one of the interference term disappears, and with the Taylor expansion of the remaining interference term, we obtain
\begin{eqnarray}
\label{ee55}
&&
Q_0 
 \approx
 p_{00} q_0 + p_{01} q_1    
 - 2 \sqrt{ p_{10}  p_{11} q_0 q_1} (p_{00} q_0 + p_{01} q_1)  \cos\theta_1,
\nonumber \\
&&
Q_1 
 \approx
 p_{10} q_0 + p_{11} q_1    
 + 2 \sqrt{ p_{10}  p_{11} q_0 q_1} (p_{00} q_0 + p_{01} q_1)  \cos\theta_1.
\end{eqnarray}
Curiously, this is another case in which single phase expression (\ref{ee24})-(\ref{ee25}) becomes valid
with the ``effective phase'' $\Theta=\Theta(\theta_1, p_{00}, p_{01})$ defined by 
\begin{eqnarray}
\label{ee56}
\cos\Theta = (p_{00} q_0 + p_{01} q_1)  \cos\theta_1.
\end{eqnarray}
In hindsight, the success of one-parameter formula \cite{FR09} could be attributed to this particular arrangement.

The current results should be compared to the value of quantum phases
obtained in our previous study of
sure-thing principle violation \cite{CT10}, in which two phases take the value $\theta_0 \sim \theta_1 \sim  2.7$ that give something rather close to the maximum destructive interference.


Our estimate of $\theta_j$ shown above  illustrates how a reasonable phenomenology can be constructed even with  experiments with lots of "unknowns".
Clearly, more experimental studies with explicit values for currently unmeasured $q_j$ and $p_{kj}$
should be very helpful.  

\section{Prospects}
We have shown in this work that it is possible to understand the two paradoxical phenomena in psychology, sure-thing principle violation and conjunction fallacy, both found by the Tversky group, in a unified single framework.  They can be
regarded as two extreme special cases of non-additivity of probabilistic decision process, for which psychological understanding are still lacking, but which are successfully describable by phenomenological formula derived from quantum decision theory.
Judging from the quantitative success of the result, it appears likely, that our approach gives a basis
for viable phenomenologies on human decision process for which theories based on classical probability have failed.  It would be interesting to combine our theory with experimental studies that utilize fMRI, with which it might even be possible  identify neural signal behind the ``quantum interference'' related  phenomena.

%
The conjunction fallacy could be studied also within the framework of nonlinear subjective probability in prospect theory \cite{KT79, TK92}. 
This theory is based on the concept of probability weighting with {\it sub-additive probability} for which probability of all events do not necessarily add up to unity.  It predicts, among other things,  the concavity alternation of $Q_k(p_{kj} q_j)$  at around $p_{kj} q_j=1/e \approx 0.37$  \cite{PR98}, and two-valued $Q_k$ at around  $p_{kj} q_j=0.5$, en route to the description of the conjunction fallacy.  With the current status of quantity and quality of the experiments, it still remains to be seen whether those effects are actually observed in the data.
Better experimental data in the future should be instrumental in sorting out the phenomenological effectiveness of both subjective probability theory and quantum theory of the conjunction fallacy.

\bigskip
This research was supported  by the Japan Ministry of Education, Culture, Sports, Science and Technology under the Grant number 21540402, and also under the
Grant- in-Aid for Scientific Research program ``Global COE''.
\bigskip

\end{document}